# SIMULATION FROM ENDPOINT-CONDITIONED, CONTINUOUS-TIME MARKOV CHAINS ON A FINITE STATE SPACE, WITH APPLICATIONS TO MOLECULAR EVOLUTION

By Asger Hobolth[1] and Eric A. Stone

*Aarhus University and North Carolina State University*

Analyses of serially-sampled data often begin with the assumption that the observations represent discrete samples from a latent continuous-time stochastic process. The continuous-time Markov chain (CTMC) is one such generative model whose popularity extends to a variety of disciplines ranging from computational finance to human genetics and genomics. A common theme among these diverse applications is the need to simulate sample paths of a CTMC conditional on realized data that is discretely observed. Here we present a general solution to this sampling problem when the CTMC is defined on a discrete and finite state space. Specifically, we consider the generation of sample paths, including intermediate states and times of transition, from a CTMC whose beginning and ending states are known across a time interval of length $T$. We first unify the literature through a discussion of the three predominant approaches: (1) modified rejection sampling, (2) direct sampling, and (3) uniformization. We then give analytical results for the complexity and efficiency of each method in terms of the instantaneous transition rate matrix $Q$ of the CTMC, its beginning and ending states, and the length of sampling time $T$. In doing so, we show that no method dominates the others across all model specifications, and we give explicit proof of which method prevails for any given $Q, T$, and endpoints. Finally, we introduce and compare three applications of CTMCs to demonstrate the pitfalls of choosing an inefficient sampler.

**1. Introduction.** This paper considers the problem of conditional sampling from a continuous-time Markov chain (CTMC) defined on a discrete and finite state space. In the ideal case, given continuously-observed sample paths, statistical inference is straightforward: the sufficient statistics

Received December 2007; revised December 2008.

[1]Supported by the US National Institute of Health (Grant R01 GM070806).

are simply the number of transitions between any two states and the total time spent in each state [e.g., Guttorp (1995), Section 3.7]. In most applications of CTMCs, however, the stochastic process $\{X(t): 0 \leq t \leq T\}$ serves as a continuous-time model for data sampled at discrete points in time $T_0 = 0 < T_1 < \cdots < T_{N-1} < T_N = T$. Sometimes this discretized problem remains amenable to closed-form analysis, and in such cases sampling is unnecessary. Examples from the literature include Holmes and Rubin (2002) and Hobolth and Jensen (2005), both of whom motivated EM algorithms by showing how to calculate the number of transitions between any two states and the time spent in a state for an endpoint-conditioned CTMC. More recently, Siepel, Pollard and Haussler (2006) showed how to calculate the probability mass function for the number of transitions, and Minin and Suchard (2008) derived analytically tractable results for the moments of the number of transitions between any two states. Our work complements these approaches, focusing on the case where parametric inference relies upon the simulation of continuous sample paths from the CTMC conditional on $X(T_0), \ldots, X(T_N)$. As a consequence of the Markov assumption, knowledge of the data $X(T_0), \ldots, X(T_N)$ effectively partitions the process into independent components $\{X(t): T_k \leq t \leq T_{k+1}\}$ whose endpoints $X(T_k)$ and $X(T_{k+1})$ are known. Thus, sampling a realization from $\{X(t): 0 \leq t \leq T\}$ given the observed data amounts to sampling from $N$ independent and identical CTMCs, each conditioned on its endpoints $X(T_k)$ and $X(T_{k+1})$ and spanning a fixed interval of time $T_{k+1} - T_k$. In what follows, we specialize to one of these $N$ components, focusing on how to simulate sample paths from a CTMC when only its endpoints are known. Crucially, while calculating the sufficient statistics from the simulated data remains easy, the simulations themselves may be prohibitively time-consuming without an efficient strategy for generating sample paths.

The wide applicability of CTMCs to serially-sampled data has accelerated their entry into the interdisciplinary literature, and several recent papers have specifically considered the problem of sampling paths from an endpoint-conditioned process. Blackwell (2003), for example, discusses a naive rejection sampling method as it applies to the analysis of radio-tracking data. A similar approach is considered in Bladt and Sørensen (2005), the motivation there coming from the field of mathematical finance. For our purposes, a naive rejection sampling method is one that simulates sample paths forward in time according to the specified process: as a consequence of endpoint conditioning, rejection of a sample path occurs whenever there is disagreement between the simulated and observed ending states. Nielsen (2002) improves upon the naive rejection sampling approach in an application to molecular evolution. His method, which we call "modified rejection sampling," conditions on the event that one state change must have occurred in cases where the observed endpoints of the process are not the same. Nevertheless, despite

Nielsen's improvement, sampling forward in time without specific regard to the ending state may lead to a prohibitively low rate of path acceptance. As an alternative, Hobolth (2008) suggests a direct sampling procedure based on analytical expressions for the probabilities of state transitions and their waiting times. A final approach, often called uniformization, originates with the work of Jensen (1953). The idea is to construct a related process that permits virtual transitions (in which the state does not change) so that the number of state transitions in a time interval can be seen as Poisson distributed and the state transitions themselves constitute a Markov chain. Sampling from this related process is equivalent to sampling from the target CTMC provided that the virtual changes are ignored. Recent applications can be found in Fearnhead and Sherlock (2006), Lartillot (2006), Mateiu and Rannala (2006), and Rodrigue, Philippe and Lartillot (2008).

Though developed for distinct applications, each of the aforementioned path-sampling procedures simulates from the same conditional distribution and thus solves the same problem. In light of this interchangeability, and because of the importance of path sampling to statistical inference on endpoint-conditioned CTMCs, it is imperative to ask whether one procedure should be preferred for its computational efficiency. The remainder of the paper seeks an exhaustive answer to this question. We first give a thorough discussion in Section 2 of the three sampling strategies: (1) modified rejection sampling, (2) direct sampling, and (3) uniformization. In Section 3 we introduce three CTMCs that highlight the heterogenous relationship between parameterization of the stochastic process and the efficiency of each sampling strategy. Section 4 then generalizes the results of Section 3 by providing analytical expressions for the efficiency of each sampler for arbitrary parameterizations of the CTMC. We conclude by summarizing the results into recommendations on how to best simulate sample paths from any given CTMC. In doing so, we also identify cases for which one or more of the potential strategies is guaranteed to fail.

**2. Sampling strategies for endpoint-conditioned chains.** In this section we review three strategies for simulating a realization of a finite-state CTMC $\{X(t) : 0 \leq t \leq T\}$ conditional on its beginning state $X(0) = a$ and ending state $X(T) = b$. The chain is defined by its instantaneous rate matrix $Q$ with off-diagonal entries $Q_{ab} \geq 0$ and diagonal entries $Q_{aa} = -\sum_{b \neq a} Q_{ab} = -Q_a < 0$. We make the futher assumption that $X(t)$ is irreducible and positive recurrent so that a stationary distribution $\pi$ exists. Finally, unless otherwise noted, we set $\sum_c \pi_c Q_c = 1$ so that one state change is expected per unit time.

To understand the sampling difficulties associated with conditioning a CTMC on its endpoints, it is useful to first review how one proceeds when the ending state $X(T)$ is unobserved. Simulating a sample path of $\{X(t) : 0 \leq$ $t \leq T$} that begins at $X(0) = a$ can be accomplished by a simple iterative procedure. The key observation is that the waiting time $\tau$ to the first state change is exponentially distributed with mean $1/Q_a$. If $\tau > T$, there is no state change in the interval $[0, T]$, and the corresponding sample path is constant; otherwise, a new state $c$ is drawn from the discrete probability distribution with probability masses $Q_{ac}/Q_a$ and the procedure is iterated for the shorter time interval $[\tau, T]$ (or, equivalently, for $[0, T - \tau]$). For reference, we present this forward sampling algorithm below:

ALGORITHM 1 (Forward sampling).

1. Sample $\tau \sim Exponential(Q_a)$. If $\tau \geq T$, we are done: $X(t) = a$ for all $t \in [0, T]$.
2. If $\tau < T$, choose a new state $c \neq a$ from a discrete probability distribution with probability masses $Q_{ac}/Q_a$. Repeat the procedure with new beginning state $c$ and new time interval $[\tau, T]$.

Under the assumption that the ending state $X(T) = b$ is observed, conditioning excludes all paths sampled from the preceding algorithm that fail to end in state $b$. This is the essence of the rejection sampling approach, whose modification by Nielsen we discuss in the next subsection.

2.1. *Rejection sampling.* As implemented in Blackwell (2003) and Bladt and Sørensen (2005), naive rejection sampling uses forward sampling to generate candidate sample paths of an endpoint-conditioned CTMC. From these, the acceptable candidates are those for which the simulated ending state and the observed ending state are the same. In particular, when sampling forward, the probability of hitting the observed ending state $b$ is $P_{ab}(T) = \exp(Qt)_{ab}$. Thus, if $T$ is large, this probability approximately equals the stationary probability $\pi_b$ of $b$. Conversely, if $T$ is small and $a \neq b$, the probability is approximately $Q_{ab}T$. It follows that in case of (i) large time $T$ and small stationary probability $\pi_b$, or (ii) different states $a \neq b$ and small time $T$, naive rejection sampling is bound to fail. Nielsen's modification improves upon naive rejection sampling in the latter case Nielsen (2002). By a conditioning argument, the time $\tau$ to the first state change given at least one state change occurs before $T$ and $X(0) = a$ has density

(2.1)     $$f(\tau) = \frac{Q_a e^{-\tau Q_a}}{1 - e^{-TQ_a}}, \qquad 0 \leq \tau \leq T.$$

The corresponding cumulative distribution function is

$$F(\tau) = \frac{1 - e^{-\tau Q_a}}{1 - e^{-TQ_a}}, \qquad 0 \leq \tau \leq T,$$

with explicit inverse

$$F^{-1}(u) = -\log[1 - u(1 - e^{-TQ_a})]/Q_a.$$

Thus, upon sampling $u$ from a $Uniform(0,1)$ distribution, transformation yields the sample waiting time $F^{-1}(u)$ to the first state change of the CTMC.

ALGORITHM 2 (Modified rejection sampling).  If $a = b$:

1. Simulate from $\{X(t) : 0 \leq t \leq T\}$ using the forward sampling algorithm.
2. Accept the simulated path if $X(T) = a$; otherwise, return to step 1 and begin anew.

If $a \neq b$:

1. Sample $\tau$ from the density (2.1) using the inverse transformation method, and choose a new state $c \neq a$ from a discrete probability distribution with probability masses $Q_{ac}/Q_a$.
2. Simulate the remainder $\{X(t) : \tau \leq t \leq T\}$ using the forward sampling algorithm from the beginning state $X(\tau) = c$.
3. Accept the simulated path if $X(T) = b$; otherwise, return to step 1 and begin anew.

In short, modified rejection sampling explicitly avoids simulating constant sample paths when it is known that at least one state change must take place. This is particularly beneficial when $T$ is small, as the naive approach will be dominated by wasted constant sample paths whose ending state remains $a$ [which occurs with probability approximately $(1 - Q_a T)$]. Nevertheless, if the transition from $a$ to $b$ is unlikely so that $Q_{ab}/Q_a$ is small, then essentially every sample path will still be rejected. In such a setting, either direct sampling or uniformization is required.

2.2. *Direct sampling.* The direct sampling procedure of Hobolth (2008) requires that the instantaneous rate matrix $Q$ admits an eigenvalue decomposition. Under that assumption, let $U$ be an orthogonal matrix with eigenvectors as columns and let $D_\lambda$ be the diagonal matrix of corresponding eigenvalues such that $Q = U D_\lambda U^{-1}$. Then, for any time $t$, the transition probability matrix of the CTMC can be calculated as

$$(2.2) \qquad P(t) = e^{Qt} = U e^{tD_\lambda} U^{-1} \quad \text{and} \quad P_{ab}(t) = \sum_j U_{aj} U_{jb}^{-1} e^{t\lambda_j}.$$

Consider first the case where the endpoints of the CTMC are identical so that $X(0) = X(T) = a$. The probability that there are no state changes in the time interval $[0, T]$ conditional on $X(0) = a$ and $X(T) = a$ is given by

$$(2.3) \qquad p_a = \frac{e^{-Q_a T}}{P_{aa}(T)}.$$

Thus, with probability $p_a$, a sample path from the CTMC will be the constant $X(t) = a$. Furthermore, with probability $(1 - p_a)$, at least one state change occurs. Thus, when $X(0) = X(T) = a$, the sample path is constant with probability $p_a$, and has at least one change with probability $(1 - p_a)$.

Next consider the case where $X(0) = a$ and $X(T) = b$, with $a \neq b$. Let $\tau$ denote the waiting time until the first state change. The conditional probability that the first state change is to $i$ at a time smaller than $t$ is

$$P(\tau \leq t, X(\tau) = i | X(0) = a, X(T) = b)$$
$$= P(\tau \leq t, X(\tau) = i, X(0) = a | X(T) = b) / P(X(0) = a | X(T) = b)$$
$$= \int_0^t Q_a e^{-Q_a z} \frac{Q_{ai}}{Q_a} \frac{P_{ib}(T - z)}{P_{ab}(T)} \, dz$$
$$= \int_0^t f_i(z) \, dz,$$

where $f_i(z)$ is the integrand. Specifically, conditional on the endpoints $X(0) = a$ and $X(0) = b$, the probability $p_i$ that the first state change is to $i$ is

$$(2.4) \qquad p_i = \int_0^T f_i(t) \, dt, \qquad i \neq a, a \neq b.$$

Using (2.2), we can rewrite the integrand as

$$(2.5) \quad f_i(t) = Q_{ai} e^{-Q_a t} \frac{P_{ib}(T - t)}{P_{ab}(T)} = \frac{Q_{ai}}{P_{ab}(T)} \sum_j U_{ij} U_{jb}^{-1} e^{T\lambda_j} e^{-t(\lambda_j + Q_a)},$$

which renders the integral in (2.4) straightforward. We get

$$(2.6) \qquad p_i = \frac{Q_{ai}}{P_{ab}(T)} \sum_j U_{ij} U_{jb}^{-1} J_{aj},$$

where

$$J_{aj} = \begin{cases} T e^{T\lambda_j}, & \text{if } \lambda_j + Q_a = 0, \\ \dfrac{e^{T\lambda_j} - e^{-Q_a T}}{\lambda_j + Q_a}, & \text{if } \lambda_j + Q_a \neq 0. \end{cases}$$

We now have a procedure for simulating the next state and the waiting time before the state change occurs. Iterating the procedure allows us to simulate a sample path $\{X(t) : 0 \leq t \leq T\}$ that begins in $X(0) = a$ and ends in $X(T) = b$.

ALGORITHM 3 (Direct sampling).

1. If $a = b$, sample $Z \sim \text{Bernoulli}(p_a)$, where $p_a$ is given by (2.3). If $Z = 1$, we are done: $X(t) = a, 0 \leq t \leq T$.

2. If $a \neq b$ or $Z = 0$, then at least one state change occurs. Calculate $p_i$ for all $i \neq a$ from (2.6). Sample $i \neq a$ from the discrete probability distribution with probability masses $p_i/p_{-a}, 0 \leq t \leq T$, where $p_{-a} = \sum_{j \neq a} p_j$. [Note that $p_{-a} = 1$ when $a = b$ and $p_{-a} = (1 - p_a)$ otherwise.]

3. Sample the waiting time $\tau$ in state $a$ according to the continuous density $f_i(t)/p_i, 0 \leq t \leq T$, where $f_i(t)$ is given by (2.5). Set $X(t) = a, 0 \leq t < \tau$.

4. Repeat procedure with new starting value $i$ and new time interval of length $T - \tau$.

REMARK 4. In step 3 above, we simulate from the scaled density (2.5) by finding the cumulative distribution function and then use the inverse transformation method. To calculate the cumulative distribution function, note that

$$\int_0^t e^{T\lambda_j} e^{-s(\lambda_j + Q_a)} \, ds = \begin{cases} t e^{T\lambda_j}, & \text{if } \lambda_j + Q_a = 0, \\ e^{T\lambda_j} \dfrac{1}{\lambda_j + Q_a}(1 - e^{-t(\lambda_j + Q_a)}), & \text{if } \lambda_j + Q_a \neq 0. \end{cases}$$

To use the inverse transformation method, we must find the time $t$ such that $F(t) - u = 0$, where $F(t)$ is the cumulative distribution function and $0 < u < 1$. In subsequent sections, we have used a (numerical) root finder for this purpose.

2.3. *Uniformization.* The final strategy that we consider permits sampling from $X(t)$ through construction of an auxilliary stochastic process $Y(t)$. Let $\mu = \max_c Q_c$ and define the process $Y(t)$ by letting the state changes be determined by a discrete-time Markov process with transition matrix

$$(2.7) \qquad\qquad R = I + \frac{1}{\mu} Q.$$

Note that, by construction, we allow *virtual* state changes in which a jump occurs but the state does not change. Indeed, virtual state changes for state $a$ are possible if $R_{aa} > 0$. Next, let the epochs of state changes be determined by an independent Poisson process with rate $\mu$. The stochastic process $Y(t)$ is called a *Markov chain subordinated to a Poisson process* and is equivalent to the original continuous-time Markov chain $X(t)$ as the following calculation shows:

$$(2.8) \qquad P(t) = e^{Qt} = e^{\mu(R-I)t} = e^{-\mu t} \sum_{n=0}^{\infty} \frac{(\mu t R)^n}{n!} = \sum_{n=0}^{\infty} e^{-\mu t} \frac{(\mu t)^n}{n!} R^n.$$

This approach is commonly referred to as uniformization, and we adopt that language here. In what follows, we describe how uniformization can be used to construct an algorithm for exact sampling from $X(t)$, conditional on the beginning and ending states.

It follows directly from (2.8) that the transition function of the Markov chain subordinated to a Poisson process is given by

$$P_{ab}(t) = P(X(t) = b|X(0) = a) = e^{-\mu t}1_{(a=b)} + \sum_{n=1}^{\infty} e^{-\mu t}\frac{(\mu t)^n}{n!}R_{ab}^n.$$

Thus, the number of state changes $N$ (including the virtual) for the conditional process that starts in $X(0) = a$ and ends in $X(T) = b$ is given by

$$(2.9) \qquad P(N = n|X(0) = a, X(T) = b) = e^{-\mu T}\frac{(\mu T)^n}{n!}R_{ab}^n/P_{ab}(T).$$

Given the number of state changes $N = n$, the times $t_1, \ldots, t_n$ at which those state changes occur are uniformly distributed in the time interval $[0, T]$. Furthermore, the state changes $X(t_1), \ldots, X(t_{n-1})$ are determined by a Markov chain with transition matrix $R$ conditional on the beginning state $X(0) = a$ and ending state $X(t_n) = b$.

Putting these things together, we have the following algorithm for simulating a continuous-time Markov chain $\{X(t) : 0 \leq t \leq T\}$ conditional on the starting state $X(0) = a$ and ending state $X(T) = b$.

ALGORITHM 5 (Uniformization).

1. Simulate the number of state changes $n$ from the distribution (2.9).
2. If the number of state changes is 0, we are done: $X(t) = a, 0 \leq t \leq T$.
3. If the number of state changes is 1 and $a = b$, we are done: $X(t) = a, 0 \leq t \leq T$.
4. If the number of state changes is 1 and $a \neq b$ simulate $t_1$ uniformly random in $[0, T]$, we are done: $X(t) = a, t < t_1$, and $X(t) = b, t_1 \leq t \leq T$.
5. When the number of state changes $n$ is at least 2, simulate $n$ independent uniform random numbers in $[0, T]$ and sort the numbers in increasing order to obtain the times of state changes $0 < t_1 < \cdots < t_n < T$. Simulate $X(t_1), \ldots, X(t_{n-1})$ from a discrete-time Markov chain with transition matrix $R$ and conditional on starting state $X(0) = a$ and ending state $X(t_n) = b$. Determine which state changes are virtual and return the remaining changes and corresponding times of change.

REMARK 6. In Step 1 above, we find the number of state changes $n$ by simulating $u$ from a Uniform$(0, 1)$ distribution and letting $n$ be the first time the cumulative sum of (2.9) exceeds $u$. When calculating the cumulative sum we need to raise $R$ to powers 1 through $n$. These powers of $R$ are stored because they are required in Step 5 of the algorithm. We use the eigenvalue decomposition (2.2) of $Q$ to calculate $P_{ab}(t)$.

REMARK 7. In Step 5 above we simulate $X(t_i), i = 1, \ldots, n-1$, from the discrete distribution with probability masses

$$P(X(t_i) = x_i | X(t_{i-1}) = x_{i-1}, X(t_n) = b) = \frac{R_{x_{i-1}, x_i}(R^{n-i})_{x_i, b}}{(R^{n-i+1})_{x_{i-1}, b}}.$$

Thus far we have outlined three competing strategies for simulating sample paths from an endpoint-conditioned CTMC. Though our discussion has been agnostic to the number of desired sample paths, this quantity has a direct and varied impact on the computational efficiency of each sampler. For example, while both direct sampling and uniformization require a possibly time-consuming eigendecomposition of $Q$, it is clear that one such computation will suffice even when multiple sample paths are desired. The number of sample paths desired from an endpoint-condtioned CTMC is application driven: estimation of some quantity, such as the expected number of visits to a given state, may require many sample paths, whereas the updating step in a Bayesian computation may require as few as one. Rather than exhaust potential applications, we have chosen to formally analyze the static and dynamic costs associated with each sampling strategy. We defer this discussion to Section 4, using the next section to demonstrate by example that no one strategy dominates the others when only one sample path is required.

**3. Comparison by example.** To illustrate the strategies detailed above, in this section we introduce three explicit examples of CTMCs for which the performance of each sampler can be directly compared. We begin with a pair of CTMCs in common use for molecular evolutionary studies; each provides a unique stochastic description of how DNA sequences evolve over time. For these and the remaining example, we compare the computational demands (measured as CPU time) of modified rejection sampling, of direct sampling, and of uniformization. For each example, the computational demands are accumulated over 100 independent samples.

In what follows, note that while rejection sampling and uniformization do not require any numerical approximations, direct sampling requires a root finder. The numerical approximation of the root finder can be made arbitrarily precise, but the choice of precision affects the running time. Without loss of generality, we have chosen the default settings of the root finder in the statistical programming language R (www.r-project.org, Version 2.0.0). The root finder typically converges in 4 to 8 iterations. The programs are run on an Intel 2.40 GHz Pentium 4 processor and are available in the supplementary material [Hobolth and Stone (2009)].

3.1. *Example 1: Molecular evolution on the nucleotide level.* We first consider a popular model of DNA sequence evolution at the nucleotide level.

The state space for a particular site in a DNA sequence is of size 4 corresponding to the DNA building blocks adenine (A), guanine (G), cytosine (C), and thymine (T). The HKY model of Hasegawa, Kishino and Yano (1985) describes the evolution of one site in a DNA sequence through an instantaneous rate matrix of the form

$$Q = (1/s) \begin{bmatrix} \cdot & \kappa\pi_{\text{G}} & \pi_{\text{C}} & \pi_{\text{T}} \\ \kappa\pi_{\text{A}} & \cdot & \pi_{\text{C}} & \pi_{\text{T}} \\ \pi_{\text{A}} & \pi_{\text{G}} & \cdot & \kappa\pi_{\text{T}} \\ \pi_{\text{A}} & \pi_{\text{G}} & \kappa\pi_{\text{C}} & \cdot \end{bmatrix},$$

where the states appear in the order A, G, C, T and the diagonal elements of $Q$ are such that the rows sum to zero. Note that state changes of the CTMC are called 'substitutions' in this context to reflect that the nucleotide in a particular site has been substituted by another. The HKY model is reversible and has stationary distribution $\pi = (\pi_{\text{A}}, \pi_{\text{G}}, \pi_{\text{C}}, \pi_{\text{T}})$. The ts/tv rate ratio parameter $\kappa$ is used to distinguish between transitions [substitutions between purines (A, G) or between pyrimidines (C, T)] and transversions (substitutions between a purine and a pyrimidine). The scaling parameter $s = s(\kappa, \pi)$ is chosen such that $\sum_{a=1}^{4} Q_a \pi_a = 1$, implying that $t$ substitutions are expected in $t$ time units. In this application, we use the parameter values $\kappa = 2$ and $\pi = (0.2, 0.3, 0.3, 0.2)$. The scaling parameter calibrates the intensity of substitutions per unit time; for context, note that the expected number of substitutions per site between humans and chimpanzees is roughly 0.01 [The Chimpanzee Sequencing and Analysis Consortium (2005)] and between humans and mice is roughly 0.50 [Mouse Genome Sequencing Consortium (2002)].

Figure 1 plots the computational demands of each sampling strategy against evolutionary distance, measured equivalently as the expected number of substitutions (CTMC state changes) or as units of time. The plot on the left demonstrates the case where the beginning and ending states are both A; by contrast, on the right the beginning state remains A while the ending state is G. The figure reveals rejection sampling to be by far the most efficient algorithm here. Moreover, direct sampling is more efficient than uniformization when the endpoints are the same and the evolutionary distance is shorter than one expected substitution per site. When the endpoints are different, uniformization is more efficient than direct sampling.

3.2. *Example 2: Molecular evolution on the codon level.* For protein-coding DNA sequences, the natural state space consists of nucleotide triplets (called codons). There are $4^3 = 64$ possible nucleotide triplets, but the three stop codons TGA, TAG, and TAA do not appear within a protein. The $64 - 3 = 61$ remaining codons constitute the state space and are called the sense codons. Each of the 61 sense codons deterministically translates into one of

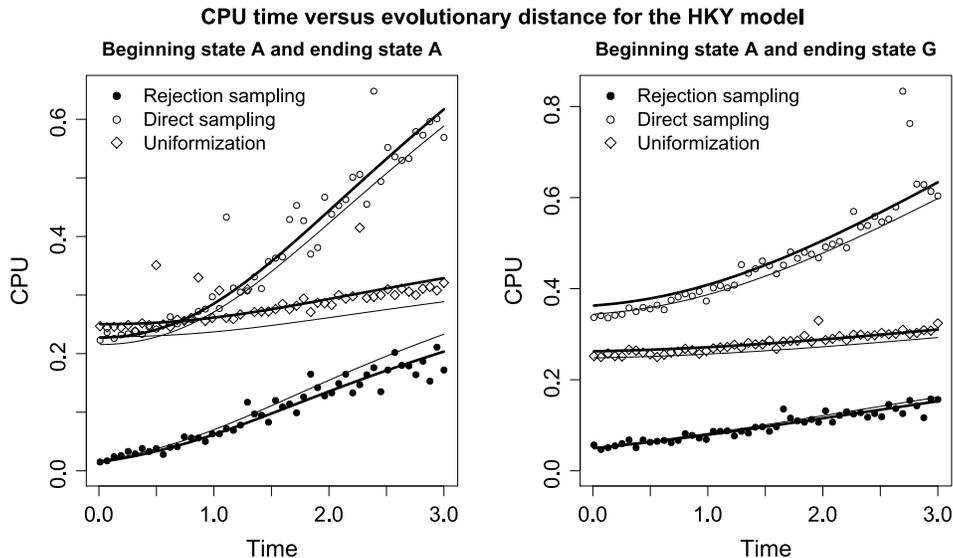

**CPU time versus evolutionary distance for the HKY model**

Fig. 1. *CPU time versus evolutionary distance for the HKY model. In both plots the beginning state is* A. *In the left plot the ending state is also* A *and in the right plot the ending state is* G. *Rejection sampling requires less CPU time than direct sampling and uniformization. The solid thick lines show predicted CPU times when the cost of initialization and recursion is fitted to the observed CPU times (see Sections 4.1–4.3). The solid thin lines show predicted CPU times when the cost of initialization and recursion is estimated from a simulation study of reversible rate matrices (see Section 4.4). Here and in Figures 2 and 3 the expected number of recursion steps was calculated analytically using the formulas in the text.*

20 amino acids, and thus distinct codons translate into the same amino acid. Substitutions between codons that translate into the same amino acid are called synonymous (or silent), while substitutions between different amino acids are called nonsynonymous (or nonsilent).

In 1994, Goldman and Yang (1994) formulated a model on the space of sense codons that is still in common use today. The GY model, a natural extension of the HKY model described above, is reversible with stationary distribution $\pi = (\pi_1, \ldots, \pi_{61})$ and incorporates a ts/tv rate ratio $\kappa$. The GY model also distinguishes between synonymous and nonsynonymous substitutions through a parameter $\omega$. The off-diagonal entries in the instantaneous GY rate matrix are given by

$$Q_{ab} = (1/s) \begin{cases} 0, & \text{if } a \text{ and } b \text{ differ at more that one position,} \\ \pi_b, & \text{for synonymous transversions,} \\ \kappa\pi_b, & \text{for synonymous transversions,} \\ \omega\pi_b, & \text{for nonsynonymous transversions,} \\ \omega\kappa\pi_b, & \text{for nonsynonymous transversions,} \end{cases}$$

where $s = s(\omega, \kappa, \pi)$ is again chosen such that $t$ substitutions are expected in $t$ time units (i.e., $\sum_a Q_a \pi_a = 1$).

In our application, we choose $\kappa = 2$ so that transitions are favored over transversions and take $\omega = 0.01$ so that synonymous changes are far more likely than nonsynonymous changes. We choose $\pi$ based on established patterns of codon usage, and note that these frequencies are quite heterogenous: the smallest entry is GGG ($\pi_{\text{GGG}} = 0.0042$) and the largest entry is GAG ($\pi_{\text{GAG}} = 0.0426$).

For these specifications, Figure 2 plots the computational demands of each sampling strategy against evolutionary distance. In each plot, the starting state is AAA. The leftmost plot compares performance when the ending state is AAG. Note that the substitution from AAA to AAG is a synonymous transition (both AAA and AAG code for the amino acid lysine) and that the frequency for AAG is 0.0396. Because synonymous transitions are very likely, the plot confirms that rejection sampling will be very efficient. Contrast this observation with the middle plot in which the ending state is AAC. The codon AAC translates to asparagine so the substitution from the beginning to the ending state is a less likely nonsynonymous transversion. This is reflected in the poor performance of the rejection sampling algorithm. Finally, the rightmost plot demonstrates what occurs when the final state is TTT. In this case, rejection sampling is not feasible because the probability of ending in the final state is effectively zero.

3.3. *Example 3: Molecular evolution on the sequence level.* The examples above seem to indicate that rejection sampling and uniformization have the most utility, but it is easy to conceive of an application for which direct sampling is most efficient. As the two previous examples show, the efficiency of rejection sampling is tied to its acceptance probability; if the observed ending state is unlikely, a large fraction of sample paths will be destined for rejection. Uniformization, on the other hand, can be inefficient when many virtual substitutions are required. With that background, we consider the extension of the HKY model described below.

Recall the HKY rate matrix (with $\nu$ instead of $\pi$)

$$Q = \begin{bmatrix} \cdot & \kappa\nu_{\text{G}} & \nu_{\text{C}} & \nu_{\text{T}} \\ \kappa\nu_{\text{A}} & \cdot & \nu_{\text{C}} & \nu_{\text{T}} \\ \nu_{\text{A}} & \nu_{\text{G}} & \cdot & \kappa\nu_{\text{T}} \\ \nu_{\text{A}} & \nu_{\text{G}} & \kappa\nu_{\text{C}} & \cdot \end{bmatrix}.$$

Jensen and Pedersen (2000) consider so-called neighbor dependent models where the instantaneous rate at a site depends on the neighbors of the site. Jensen and Pedersen (2000) are particularly interested in CG avoidance where the rate away from C is particularly high if its right neighbor is a G. Such a model implies CG deficiency in a single sequence, which is an

### CPU time versus evolutionary distance for the GY model with beginning state AAA and various ending states

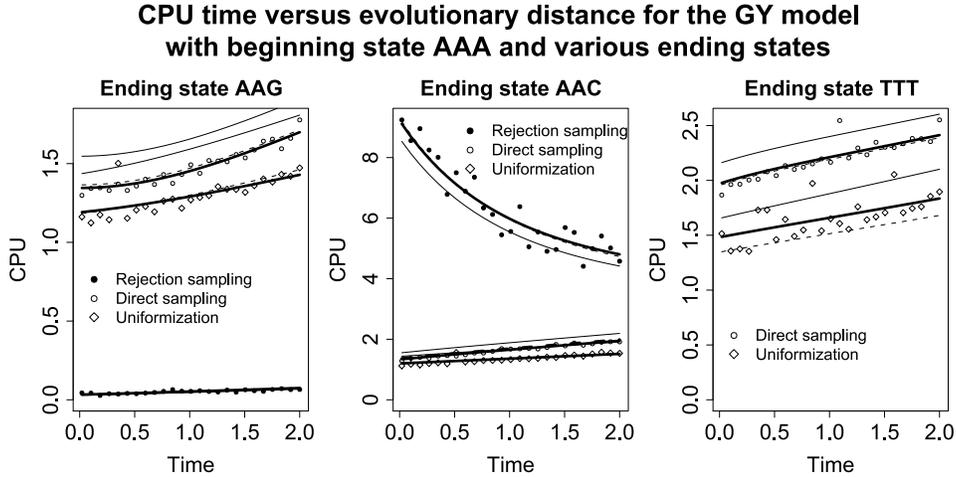

FIG. 2. *CPU time versus evolutionary distance for the GY model. In all plots the beginning state is* AAA. *In the left plot the ending state is* AAG, *in the middle plot the ending state is* AAC, *and in the right plot the ending state is* TTT. *Rejection sampling is most efficient in the situation depicted on the left, but it enters an infinite while loop on the right (and is therefore not shown). Direct sampling and uniformization have similar running times, with uniformization being slightly faster. The solid thick lines show predicted CPU times when the cost of initialization and recursion is fitted to the observed CPU times (see Sections 4.1–4.3). The solid thin lines show predicted CPU times when the cost of initialization and recursion is estimated from a simulation study of reversible rate matrices (see Section 4.4). Finally, the dashed lines show predicted CPU times when the initialization and recursion costs are estimated from a simulation study of sparse rate matrices (see Section 4.4).*

often observed phenomenon for mammalian sequences due to the process of CpG methylation-deamination. Neighbor-dependent nucleotide models are also considered in Hwang and Green (2004) and Hobolth (2008). In these two papers, a Gibbs sampling scheme is used to estimate the parameters of the model while taking the uncertainty of the neighbors into account. In particular, each single site is updated conditionally on the current values of the complete evolutionary history of the neighboring nucleotides.

Consider the evolution at a single site and assume for simplicity that the evolutionary history of the left neighbor is never a C, and the evolutionary history of the right neighbor is always a G. In this situation, a CG dinucleotide is present when the site that we consider is a C. Jensen and Pedersen (2000) model the CpG effect through increasing the rate away from CG nucleotides by multiplying each entry in the HKY rate matrix (3.3) corresponding to C with a parameter $\gamma > 1$. When the left neighbor is not a C and the right neighbor is a `G`, the rate matrix thus becomes

$$(3.1) \qquad Q^{\texttt{HKY+CG}} = \begin{bmatrix} \cdot & \kappa\nu_{\texttt{G}} & \nu_{\texttt{C}} & \nu_{\texttt{T}} \\ \kappa\nu_{\texttt{A}} & \cdot & \nu_{\texttt{C}} & \nu_{\texttt{T}} \\ \gamma\nu_{\texttt{A}} & \gamma\nu_{\texttt{G}} & \cdot & \gamma\kappa\nu_{\texttt{T}} \\ \nu_{\texttt{A}} & \nu_{\texttt{G}} & \kappa\nu_{\texttt{C}} & \cdot \end{bmatrix}.$$

The stationary distribution $\pi$ of $Q^{\texttt{HKY+CpG}}$ is given by

$$(\pi_{\texttt{A}}, \pi_{\texttt{G}}, \pi_{\texttt{C}}, \pi_{\texttt{T}}) = (\nu_{\texttt{A}}, \nu_{\texttt{G}}, \nu_{\texttt{C}}/\gamma, \nu_{\texttt{T}})/(\nu_{\texttt{A}} + \nu_{\texttt{G}} + \nu_{\texttt{C}}/\gamma + \nu_{\texttt{T}}).$$

If the parameters are $(\nu_{\texttt{A}}, \nu_{\texttt{G}}, \nu_{\texttt{C}}, \nu_{\texttt{T}}) = (0.3, 0.3, 0.2, 0.2)$ and $\gamma = 20$, we obtain the stationary distribution $(\pi_{\texttt{A}}, \pi_{\texttt{G}}, \pi_{\texttt{C}}, \pi_{\texttt{T}}) = (0.3, 0.3, 0.01, 0.2)/0.81$. Note that the stationary probability of a `C` nucleotide is now $0.01/0.81 = 0.012$.

The left-hand plot of Figure 3 illustrates the performance of the three samplers when the CTMC begins in `T` and ends in state `C`. Here the most efficient sampler depends on the time between the states: if $T < 0.3$, rejection sampling is the most efficient, if $0.3 < T < 0.9$, uniformization is the most efficient, and if $T > 0.9$, direct sampling is the most efficient. For large times,

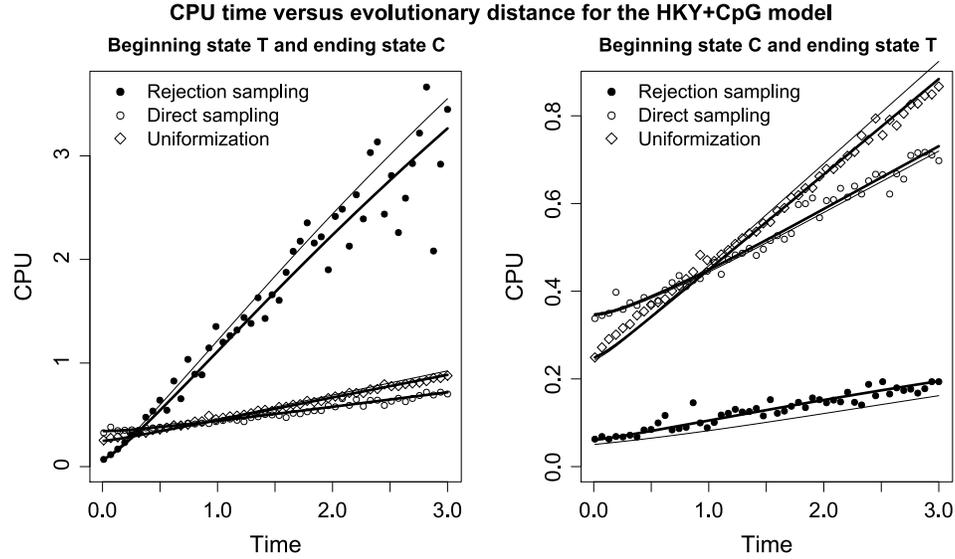

**CPU time versus evolutionary distance for the HKY+CpG model**

Fig. 3. *CPU usage versus time for the HKY + CpG rate matrix (3.1). In the left plot, the beginning state is* T *and ending state is* C. *In the right plot, the beginning state is* C *and ending state is* T. *Rejection sampling is very fast in the situation depicted on the right, but it is slow for large evolutionary distances on the left. Direct sampling and uniformization have similar running times, but direct sampling is faster for large evolutionary distances. The solid thick lines show predicted CPU times when the cost of initialization and recursion is fitted to the observed CPU times (see Sections 4.1–4.3). The solid thin lines show predicted CPU times when the cost of initialization and recursion is estimated from a simulation study of reversible rate matrices (see Section 4.4).*

rejection sampling is inefficient because it is unlikely to end in state `C`, and direct sampling becomes more efficient than uniformization because many virtual changes are required in the uniformization procedure. The right-hand plot of Figure 3 shows the case when the beginning state is `C` and the ending state is `T`. Under this scenario, rejection sampling is the most efficient sampling algorithm because the acceptance probability is high.

## 4. Complexity of samplers.

The examples in the previous section were chosen to demonstrate the heterogenous dependence of each sampling strategy upon the characteristics of the endpoint-conditioned CTMC. In particular, efficiency was shown to be impacted by each aspect of the process: the instantaneous rate matrix $Q$, the sampling time $T$, and the beginning and ending states $a$ and $b$. This section translates the qualitative observations above into quantitative proof of which sampler will be most efficient for any specification of CTMC. To accomplish this, we rely on the algorithmic descriptions of the three sampling strategies as given in Section 2. Note that the algorithms are schematically consistent, with each progressing through (1) initialization, (2) recursion, and (3) termination. Our approach is to define the fixed computational costs for the initialization and recursion steps, which we call $\alpha$ and $\beta$, respectively. As shown in Section 2, the number of recursion steps required to generate an entire sample path is stochastic, and we capture this in a random variable $L$. Thus, the computation cost of generating one sample path is

$$\alpha + \beta L$$

and the mean cost is obviously $\alpha + \beta E[L]$. In the case of rejection sampling, note that only a certain fraction of the generated sample paths will be consistent with the observed ending state and hence accepted. Ultimately, the results of this section demonstrate our ability to accurately predict the CPU time needed to produce one valid sample path from an endpoint-conditioned process. Such analysis is of great practical importance, as it allows the researcher to choose the most efficient sampler in advance.

The rest of this section is organized as follows. In Sections 4.1–4.3 we discuss complexity and derive the mean number of recursions $E[L]$ for each sampler. In Section 4.4 we first demonstrate that $\alpha$ and $\beta$ can be estimated from the size of the state space and structure of the rate matrix only. Thus, determining the values of $\alpha$ and $\beta$ is a one-time calculation. Second, we provide a strategy for choosing the most efficient sampler. The strategy depends on the mean number of recursions and estimated values of $\alpha$ and $\beta$. In case of rejection sampling, the strategy also depends on the acceptance probability. Third, we give further insight into the sampling strategies by analyzing the proposed strategy in detail for moderately large time intervals.

4.1. *Rejection sampling complexity.* Let $p_{acc}$ be the acceptance probability for the rejection sampling algorithm first described in Section 2.1. Then the expected number of samples before acceptance is its reciprocal $1/p_{acc}$. In the notation described above, the mean CPU time required to simulate one sample path is thus

$$(4.1) \qquad\qquad (\alpha + \beta E[L])/p_{acc}.$$

When the beginning and ending states take the same value, say, $a$, the acceptance probability is simply $P_{aa}(T)$. In particular, for small $T$ we have $p_{acc} \approx (1 - Q_a T)$, and for large $T$ we have $p_{acc} \approx \pi_a$. Furthermore, the expected number of recursion steps required to generate one sample path is given by

$$(4.2) \qquad\qquad E[L] = \sum_i \sum_{j \neq i} E[N_{ij}(T)|X(0) = a],$$

where $N_{ij}(T)$ is the number of state changes from $i$ to $j$ in the time interval $[0, T]$. This expectation is given by [e.g., Proposition 3.6 of Guttorp (1995)]

$$E[N_{ij}(T)|X(0) = a] = Q_{ij} \int_0^T P(X(t) = i|X(0) = a)\, dt.$$

Analytical expressions for the integral can be found by appealing to an eigendecomposition of $Q$ (see Section 2.2).

Figure 4 provides an illustration of the above considerations using the HKY model from Section 3.1 as an example. The top panel of the figure details the case when the beginning and ending states are the same (specifically, the nucleotide A). From the left, the first column plots the acceptance probability $\exp(QT)_{\text{AA}}$ against the time $T$, showing a nonlinear decrease from $p_{acc} \approx 1$ when $T$ is small to $p_{acc} \approx \pi_{\text{A}}$ when $T$ is large. The sloped dashed line plots the first-order Taylor approximation $1 - Q_{\text{A}}T$ of $p_{acc}$ that is valid for small $T$; the horizontal dashed line indicates the stationary probability $\pi_{\text{A}}$ that is the limit of $p_{acc}$ when $T$ grows large.

The second column plots CPU time spent on initialization against $T$ for a collection of simulated sample paths. A linear regression was used to estimate the initialization cost. More specifically, we generated 500 independent samples from the modified rejection sampler and recorded the time spent on initialization and recursion, respectively. The CPU time spent on initialization is proportional to $1/p_{acc}$ [recall (4.1)]; we estimated $\alpha$ using linear regression and obtained $\hat{\alpha} = 0.0509$. The third column shows the expected number of state changes $E[L]$, calculated from (4.2), as a function of time. In the fourth column we show the CPU time spent on sampling. The CPU time spent on sampling is proportional to $E[L]/p_{acc}$ [recall (4.1)]; we estimated $\beta$ using linear regression and obtained $\hat{\beta} = 0.0365$. Adding the CPU time spent on initialization and sampling gives the total CPU time spent on producing a sample path. The total CPU time and predicted CPU time is shown in the left plot of Figure 1.

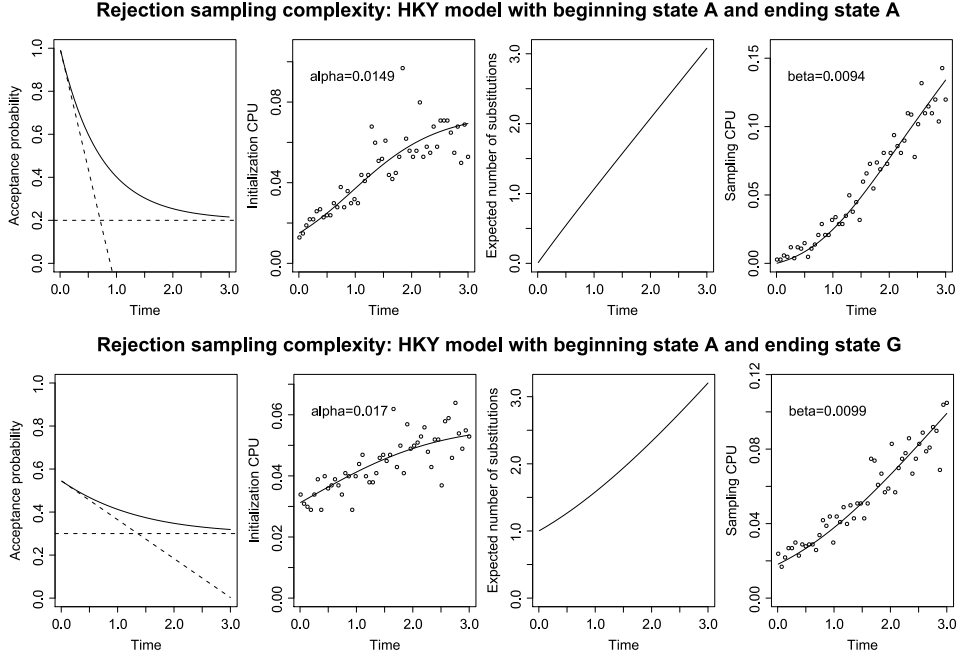

Fig. 4. *Summary statistics for the HKY model from Section 3.1. The top row shows the case when the beginning state is A and the ending state is A. In the bottom row, the beginning state is A and the ending state is G. The first column shows the probability of ending in the correct state (the acceptance probability), and the second column shows the CPU time spent on initialization. The third column shows the expected number of recursions required in each forward sample, and the fourth column shows the CPU time spent on sampling. Summing the CPU times spent on initialization and on sampling gives the total CPU time spent to produce a sample path. This total time is shown in the left-hand plot of Figure 1.*

When the beginning and ending states are different, the calculations are only slightly more complicated. To compute the acceptance probability in the case $a \neq b$, let $N(t)$ be the number of state changes of $X(t)$ in the interval $[0, t]$. We have

$$
\begin{aligned}
P_{ab}(T) &= \Pr(X(T) = b | X(0) = a) \\
&= \Pr(X(T) = b, N(T) > 0 | X(0) = a) \\
&= \Pr(X(T) = b | N(T) > 0, X(0) = a) \Pr(N(T) > 0) \\
&= p_{acc} \Pr(N(T) > 0),
\end{aligned}
$$

(4.3)

from which it is clear that

$$(4.4) \qquad p_{acc} = \frac{P_{ab}(T)}{1 - \Pr(N(T) = 0)} = \frac{P_{ab}(T)}{1 - e^{-TQ_a}}.$$

For small $T$ we have the first-order approximation

$$(4.5) \qquad p_{acc} \approx \frac{Q_{ab}}{Q_a}\Big(1 - Q_b\frac{T}{2}\Big) + \sum_{i \neq (a,b)} \frac{Q_{ai}}{Q_a}\frac{T}{2}Q_{ib},$$

and for large $T$ it is clear that $p_{acc} \approx \pi_b$.

Next we consider the number of recursion steps $L$. We know that the number of state changes is at least one because we have assumed that the beginning and ending states $a$ and $b$ are different. The probability of the first change being to state $k$ $(k \neq a)$ is $Q_{ak}/Q_a$, and the density of the time to this change is given by (2.1). Let the number of state changes from $i$ to $j(j \neq i)$ when the first substitution is to $k$ be denoted $N_{ij,k}$. The expected number of such changes in a time interval $[0, T]$ is given by

$$E[N_{ij,k}(T)] = \int_{t=0}^{T} \frac{Q_a e^{-tQ_a}}{1 - e^{-TQ_a}}\frac{Q_{ak}}{Q_a}Q_{ij}\int_{s=t}^{T} P_{ki}(s)\,ds\,dt.$$

Again, this integral can be calculated analytically using an eigenvalue decomposition of $Q$. The expected value of $L$ is given by

$$E[L] = 1 + \sum_i \sum_{j \neq i} \sum_{k \neq a} E[N_{ij,k}(T)].$$

The bottom row of Figure 4 mirrors the top row, except that here the ending state G has been chosen to be distinct from the beginning state A. As before, the first plot from the left shows the acceptance probability (4.4) against the time $T$. The sloped dashed line now shows the linear approximation (4.5), while the horizontal dashed line indicates the stationary probability $\pi_G$ of the ending state G. In the second plot, the CPU time spent on initialization is explained by the reciprocal of the acceptance probability. In the third plot, we show the expected number of substitutions, and in the last plot the CPU time spent on sampling is explained by the expected number of substitutions divided by the acceptance probability. The regression coefficient for initialization is 0.0509 and for sampling 0.0366. Note that these coefficients are very similar to what was observed in the case of equal beginning and ending states.

To complement the observations of Figure 4, recall the GY model introduced in Section 3.2. The three plots in Figure 5 mirror those in Figure 2, with each showing how the acceptance probability scales with time in the previously depicted scenario. In all cases, the beginning state is the codon AAA: from the left, the first plot considers the ending state AAG (a synonymous transition away from `AAA`), the second plot considers the ending state `AAC` (a nonsynonymous transversion away from `AAA`), and the third plot considers the ending state `TTT` (a minimum of three state changes away from `AAA`). In the first case, the acceptance probability is high and rejection sampling is efficient. In the second case, the acceptance probability is low, particularly for small $T$, and rejection sampling is less efficient. In the third case, the probability of ending up in the desired state `TTT` is smaller than $1/10^6$, and rejection sampling cannot be used. In this final case, one must use direct sampling or uniformization.

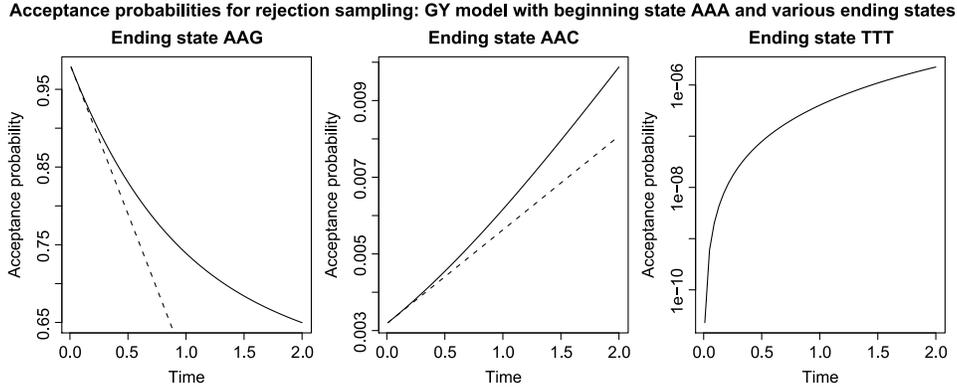

Fig. 5. *Acceptance probabilities for the GY model from Section 3.2. In all cases the beginning state is `AAA`. In the left-hand plot, the ending state is `AAG`, in the middle plot the ending state is `AAC`, and in the right-hand plot the ending state is `TTT`. Rejection sampling is very efficient in the situation depicted on the left, less efficient in the middle, and not practical in the right.*

4.2. *Direct sampling complexity.* The computational costs for direct sampling are dependent upon its initialization and the CPU time spent on sampling a new state and its corresponding waiting time. As before, the cost of generating one sample path can be written as

$$\alpha + \beta L,$$

but the initial cost $\alpha$ is much more expensive than for rejection sampling because an eigendecomposition of $Q$ is required. The expected number of recursion steps is equivalent to the number of state changes $N$ and can be found through

$$E[L] = E[N(T)|X(0) = a, X(T) = b]$$
$$= \sum_i \sum_{j \neq i} E[N_{ij}(T)|X(0) = a, X(T) = b],$$

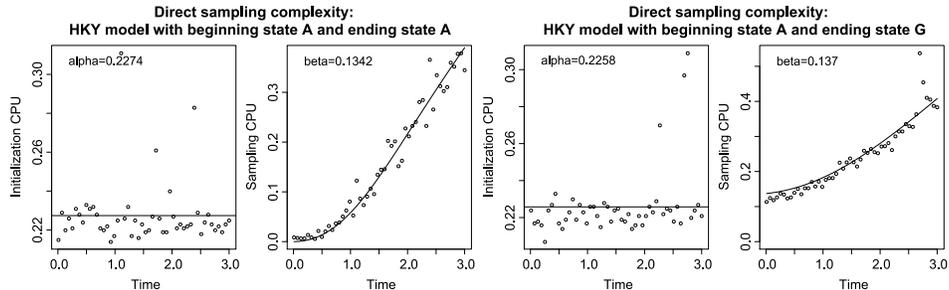

Fɪɢ. 6. *CPU time spent on direct sampling in the HKY model from Section 3.1. Two cases are considered: beginning state* A *and ending state* A *(first two plots from the left) and beginning state* A *but ending state* G *(last two plots). In both cases, the initialization CPU is constant (first and third plot). The sampling CPU is proportional to the expected number of substitutions (second and fourth plot).*

where $N_{ij}(T)$ is the number of state changes from $i$ to $j$ in the time interval $[0, T]$. These expectations can be calculated using formulas in Hobolth and Jensen (2005).

In Figure 6 we illustrate the above considerations for the HKY model from Section 3.1. The initialization cost is constant, and the number of state changes explains the cost of sampling. The initialization cost $\alpha$ is around 0.85, which is much larger than the 0.05 observed when doing rejection sampling. Moreover, the cost $\beta$ for each recursion step is 0.56, as compared to 0.04 for rejection sampling. This may seem an unfavorable comparison, but recall that rejection sampling does not guarantee that the endpoint conditions are met by its generated sample paths; if the probability of acceptance $p_{acc}$ is low, then the cost of rejection sampling given by (4.1) will be dominated by $1/p_{acc}$.

This illustrates the tradeoff that distinguishes rejection sampling from the two remaining approaches: the computational costs of rejection sampling are comparatively inexpensive, but only a fraction of the simulated sample paths from that method will be viable.

4.3. *Uniformization complexity.* The computational costs for uniformization are similar in structure to those of direct sampling. Initialization requires an eigendecomposition of $Q$ and construction of the auxiliary transition matrix $R$ in order to carry out Step 1 of the algorithm (recall Remark 6). Each recursion step consists of sampling a new state and its corresponding waiting time; examination of the uniformization algorithm reveals that the number of recursion steps $L$ is equal to the number of state changes $N(T)$ accumulated by the auxiliary chain. Thus,

$$E[L] = E[N(T)|X(0) = a, X(T) = b]$$

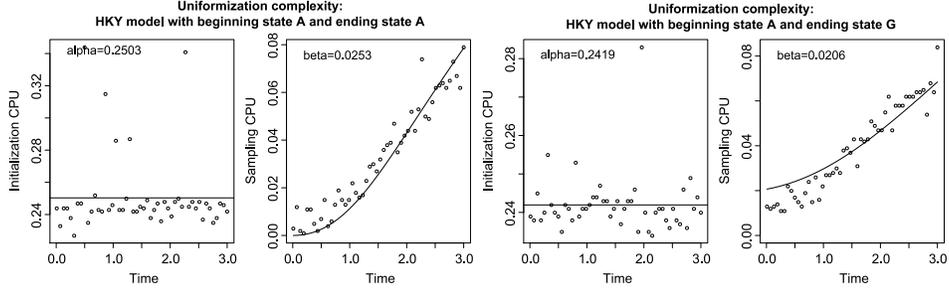

Fig. 7. *CPU time spent on uniformization in the HKY model from Section 3.1. Two cases are considered: beginning state A and ending state A (first two plots) and beginning state A but ending state G (last two plots). In both cases, the initialization CPU time is constant (first and third plot). The sampling CPU time is proportional to the expected number of substitutions (second and fourth plot).*

$$= \frac{1}{P_{ab}(T)} \sum_{n=0}^{\infty} n e^{-\mu T} \frac{(\mu T)^n}{n!} (R^n)_{ab}$$

$$= \frac{1}{P_{ab}(T)} \mu T (R e^{QT})_{ab}$$

$$= \frac{1}{P_{ab}(T)} \mu T \sum_{c} R_{ac} P_{cb}(T).$$

In particular, when $T$ is large we get $E[L] \approx \mu T$.

Figure 7 illustrates the above considerations for the HKY model from Section 3.1. As with direct sampling, the initialization cost is constant and the number of state changes (both real and virtual) explains the cost of sampling. We find $\hat{\alpha} = 1.05$, which is about the same magnitude as the initialization cost for direct sampling. In uniform sampling, the recursion step is immediate if we enter Steps 2–4. Each recursion in Step 5 is also very fast because we just have to simulate from a discrete-state Markov chain with transition probability matrix $R$ given the endpoints, and where all the relevant powers of the transition matrix $R$ are already calculated. The recursion cost $\hat{\beta} = 0.09$ is around $1/6$ of the recursion cost for direct sampling and twice as much as the recursion cost of rejection sampling.

### 4.4. *Comparison and recommendation.*

#### 4.4.1. *Comparison and recommendation for general $T$.* The preceding results explicitly relate the computational complexity of each sampling strategy to characteristics of the CTMC. This permits the three strategies to be compared to each other, but only after reliable values for $\alpha$ and $\beta$ have been obtained. We also note that the values of $\alpha$ and $\beta$ depend on the choice of

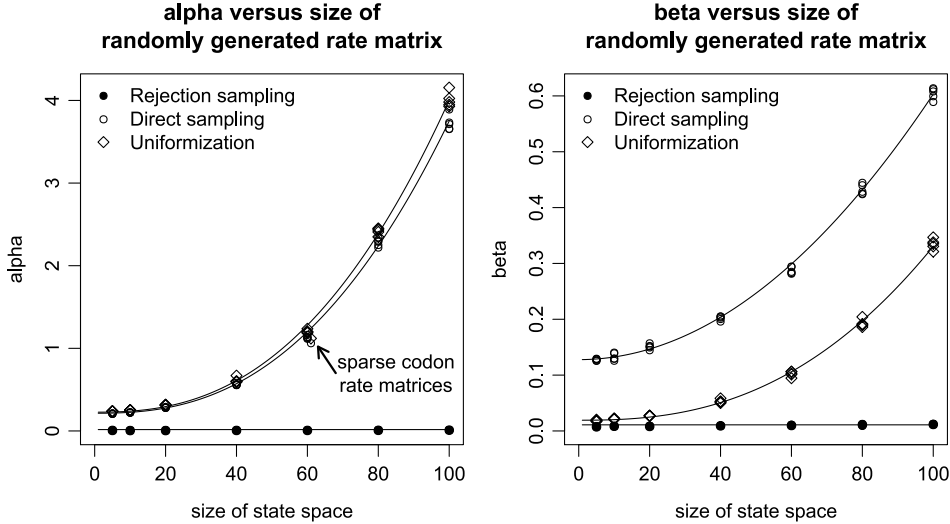

computer language. Running simulations as we have to estimate these parameters is not practical, as it compromises the gains in choosing an efficient sampler. For that reason, and to establish the generality of our observations, we sought to relate $\alpha$ and $\beta$ to the size of the state space of the CTMC.

For each of the three sampling algorithms, we estimated the values of $\alpha$ and $\beta$ for randomly simulated reversible rate matrices. Specifically, we first generated a symmetric matrix $S$ with randomly generated exponentially distributed off-diagonal entries $S_{ij} \sim \text{Exp}(1)$, $i > j$. Second, we generated the stationary distribution $\pi$ of the CTMC by sampling from a Dirichlet distribution $\text{Dir}(\alpha)$ with $\alpha = (1, \ldots, 1)$, that is, a vector of ones. The off-diagonal entries in the rate matrix become $Q_{ij} = S_{ij}\pi_j, i > j$, and $Q_{ij} = S_{ji}\pi_i, i < j$. We considered seven different state space sizes $(5, 10, 20, 40, 60, 80, 100)$, and repeated each of the simulations five times. The results are summarized in Figure 8.

Figure 8 supports our previous observations on the three sampling strategies. The costs of initialization, as quantified by $\alpha$, are as expected, with direct sampling and uniformization slowed relative to rejection sampling by their dependence on an eigenvalue decomposition of the rate matrix $Q$. Indeed, whereas the initialization cost of rejection sampling remains essentially unchanged as the state space grows, the other methods increase in runtime nonlinearly. Theory suggests that the eigenvalue decomposition that dominates direct sampling and uniformization should depend on the cube of the size of the state space, and we find in our limited sample that the relationship is best explained by an exponent of 2.5. In any case, direct sampling and uniformization are comparable in their initialization costs, with uniformization always slightly slower because it requires powers of the transition probability matrix $R$ to also be calculated. By contrast, for the state space sizes that we considered, uniformization has a substantially smaller value of $\beta$. Compare these observations to the results shown in Figures 4, 6, and 7 for the HKY model: what was true in that example with a state space of size four appears to hold consistently across our sample of simulated rate matrices of various sizes.

The most encouraging feature of Figure 8 is the apparent ease with which $\alpha$ and $\beta$ can be predicted solely from the size of the state space. As illustrated in the figure, we fit six simple models to predict $\alpha$ and $\beta$ for each of the three sampling strategies from the state space size alone. The utility of these predictions becomes clear in reference to the four-state HKY model. Previously we found $\alpha$ and $\beta$ by simulating endpoint-conditioned samples from the model of interest itself. For rejection sampling, these values were $\hat{\alpha} = 0.0149$ and $\hat{\beta} = 0.0094$ (Figure 4), which we compare to $\alpha = 0.0165$ and $\beta = 0.0109$ predicted for a state space of size four. Similarly, as shown in Figure 6, for direct sampling $(\hat{\alpha}, \hat{\beta})$ ranges from $(0.2274, 0.1342)$, when the beginning and ending states were the same, to $(0.2258, 0.1370)$, when they were different. By comparison, the predicted values for direct complexity are $\alpha = 0.2155$ and $\beta = 0.1285$. Last, recall from Figure 7 that for uniformization $(\hat{\alpha}, \hat{\beta})$ ranges from $(0.2503, 0.0253)$ for identical beginning and ending states to $(0.2419, 0.0206)$ when different. These values agree with our predictions of $\alpha = 0.2286$ and $\beta = 0.0143$, although the predicted value of $\beta$ is somewhat lower than the fit from the model itself.

It should be emphasized that the goal here is not to perfectly predict $\alpha$ and $\beta$ for any particular CTMC. Rather, the purpose of the simple models illustrated in Figure 8 is to obtain values accurate enough to decide which sampling strategy will be most efficient. With that in mind, reconsider the CPU times observed for the three examples in Section 3. In Figures 1–3 respectively, CPU time estimates for the HKY, GY, and HKY + CpG models are shown as thin lines obtained from the new predictions of $\alpha$ and $\beta$. It is clear that a practitioner, armed with only these predictions derived from the state space size, would choose the most efficient sampler in each example for virtually any combination of time and specific endpoint conditions. On the other hand, despite that success, the CPU times we predict are not uniformly accurate; the predictions for the GY model shown in Figure 2, for example, do not fit the observed data well at all. We speculated that this observed lack of fit might be a result of the structure of the GY model being too different from that of the randomly generated rate matrices used to establish our models. To pursue this hypothesis, we generated $61 \times 61$ rate matrices with the same structure as the GY model. In particular, the only nonzero off-diagonal entries in the random rate matrix are those entries where the two codons are exactly one point substitutions away from eacy other. These sparse rate matrices result in smaller values for $\alpha$ in the cases of direct sampling and uniformization, as indicated in Figure 8. As expected from Figure 2, the value of $\alpha$ for rejection sampling and $\beta$ values for all samplers are largely unaffected. The predicted CPU times from the new values of $\alpha$ and $\beta$ are shown as dashed lines and present a satisfactory fit. Thus, while our models for $\alpha$ and $\beta$ appear to be of sufficent quality to guide the correct choice of sampler, it is clear that differently structured rate matrices of the same size may yield substantially different values of $\alpha$ and $\beta$ and hence CPU times.

The preceding discussion motivates the following guidelines for choosing the most efficient sampling strategy:

1. Estimate $\alpha$ and $\beta$ for the three sampling strategies. As discussed, $\alpha$ and $\beta$ can be estimated reliably from the size of the rate matrix, allowing for some variability due to its structure.
2. Predict CPU times for rejection sampling $(\alpha_R + \beta_R E_R[L])/p_{acc}$, for direct sampling $(\alpha_D + \beta_D E_D[L])$, and for uniformization $(\alpha_U + \beta_U E_U[L])$.
3. Choose the sampler with the lowest predicted CPU time.

4.4.2. *Comparison for moderately large $T$.* We end the comparison by considering the special case when $T$ is at least moderately large. In this case some useful rules of thumb emerge. To begin, note that the expected number of iterations required for rejection sampling and for direct sampling should be approximately equal. For moderately large $T$, we can make the substitution of $(\sum_c \pi_c Q_c)T$ for $E[L]$, or just $T$, provided that the chain has been calibrated such that $\sum_c \pi_c Q_c = 1$. For uniformization, $E[L]$ is larger because of virtual state changes; under the same assumptions, here $E[L]$ can be roughly approximated by $\mu T = (\max_c Q_c)T$. Under these assumptions, virtual changes increase the number of iterations required in uniformization by a factor of

$$\nu = \max_c Q_c \Big/ \left(\sum_c \pi_c Q_c\right) = \max_c Q_c,$$

again assuming that the chain has been calibrated. In other words, the inflation factor $\nu$ is the ratio of the maximum diagonal entry of the rate matrix $Q$ to its (weighted) average diagonal entry. We obtain $\nu = (1.12, 2.22, 16.2)$ for the rate matrices in Examples 1–3, respectively. As expected, the inflation factor is very high for the rate matrix in Example 3 and explains the observations in Figure 3.

Finally, with $E[L] = T$ for both rejection sampling and direct sampling, the approximate complexities can be expressed as follows:

(4.6)

| Rejection sampling | Direct sampling | Uniformization. |
|---|---|---|
| $(\alpha_R + \beta_R T)/p_{acc}$ | $\alpha_D + \beta_D T$ | $\alpha_U + \beta_U T \nu$ |

To see the utility of these formulas, recall the results for the HKY model shown in Figure 1 and consider the moderately large time $T = 2$. Noting that for direct sampling and uniformization our approximations are not endpoint-dependent, and using estimates of $(0.2155, 0.1285)$ for $(\alpha, \beta)$ for direct sampling and $(0.2286, 0.0143)$ for uniformization, the formulas predict their CPU times to be 0.472 and 0.261, respectively, in both panels of Figure 1. Rejection sampling, of course, is dependent on the ending state, and thus the complexities for that method illustrated in the left and right plots differ. In this case, the difference is subtle because (1) the chain is nearly mixed and (2) the stationary probabilities that govern the acceptance probabilities are similar. Using 0.2 and 0.3 as the respective acceptance probabilities for A and G, and $(0.016, 0.010)$ for $(\alpha, \beta)$, we obtain 0.19 for the left plot and 0.13 for the right plot. Inspection of Figure 1 gives validity to our approximations, showing all of the predictions to be highly accurate. For the HKY + CpG model in Figure 3 we can make similar predictions. We predict the CPU times for direct sampling and uniformization to be 0.472 and 0.693, again in good agreement with both figures. Using the stationary probabilities 0.012 and 0.246 for C and T, we obtain 3.112 for the left plot and 0.155 for the right plot. These predictions are again very accurate.

In the particular case of moderately large $T$, the guideline for choosing the most efficient sampling strategy can be made even more explicit. It follows immediately from (4.6) that uniformization is more efficient than direct sampling if

$$\nu < \frac{\alpha_D + \beta_D T - \alpha_U}{\beta_U T} = \nu_{\text{critical}}.$$

Recall the transition matrix (2.7) of the auxiliary process. It is evident that if the inflation factor $\nu$ is large, then the transition matrix has one or more states where virtual state changes are very likely. In the uniformization sampling procedure, these virtual state changes have to be simulated, although they are eliminated in the final sample path. Many invisible virtual jumps thus makes uniformization less efficient. For the state space of size 4 with $(\alpha_D, \beta_D, \alpha_U, \beta_U) = (0.2155, 0.1285, 0.2286, 0.0143)$, we obtain $\nu_{\text{critical}} = 8.5$. For the HKY model we have $\nu_{\text{HKY}} = 1.12$ and for the HKY + CpG model we have $\nu_{\text{HKY+CpG}} = 16.2$, and, thus, we predict uniformization to be more efficient for large $T$ for the HKY model, while direct sampling is more efficient for the HKY + CpG model.

Similarly, it follows from (4.6) that rejection sampling is more efficient than uniformization if

$$(4.7) \qquad p_{acc} > \frac{\alpha_R + \beta_R T}{\alpha_U + \beta_U T \nu} = p_{\text{critical}}^U,$$

and more efficient than direct sampling if

$$(4.8) \qquad p_{acc} > \frac{\alpha_R + \beta_R T}{\alpha_D + \beta_D T} = p_{\text{critical}}^D.$$

For the HKY model, we get $p_{\text{critical}}^U = 0.147$. In the case of beginning state A, ending state A and for $T = 2$, we get from (4.4) that $p_{acc} = 0.254$. If the beginning state is G and $T = 2$, ending state is G and $T = 2$, we get $p_{acc} = 0.347$. Both acceptance probabilities are larger than, and we predict correctly (recall Figure 1) that rejection sampling is the most efficient algorithm in both cases.

For the HKY + CpG model we get $p_{\text{critical}}^D = 0.081$. In the case where the beginning state is T, ending state is C and $T = 2$, we obtain $p_{acc} = 0.017$, and with beginning state C, ending state T and $T = 2$, we get $p_{acc} = 0.272$. We thus correctly predict (recall Figure 3) that direct sampling is the most efficient algorithm in the first situation, while rejection sampling is more efficient in the second situation.

The approximations in (4.6) are less precise for the GY model because the larger state space increases the dependency of the beginning and ending states. However, we still get reliable predictions when applying the moderately large $T$ approximations. The predicted values in Figure 8 for the sparse codon rate matrices size 61 are $(\alpha_R, \beta_R, \alpha_D, \beta_D, \alpha_U, \beta_U) = (0.017, 0.011, 1.072, 0.305, 1.124, 0.105)$. For $T = 2$ we get $\nu_{\text{critical}} = 2.66$ and since $\nu_{\text{GY}} = 2.22$, we correctly predict uniformization to be more efficient than direct sampling. We get $p_{\text{critical}}^U = 0.024$, meaning that uniformization is also more efficient than rejection sampling if the acceptance probability is smaller than 2.4%. With $T = 2$ and beginning state AAA, we get acceptance probabilities 0.65, 0.01, and $1/10^5$ for ending states AAG, AAC, and TTT, respectively (recall Figure 5). We thus correctly predict rejection sampling to be faster than uniformization when the ending state is AAG, and slower when the ending state is AAC or TTT.

To summarize, this section shows that when the cost of initialization $\alpha$ and the cost of a recursion step $\beta$ are known, we can accurately predict the time it takes to produce a single sample from any of the three simulation procedures. We have thus demonstrated that choosing among the simulation procedures is an objective task that can be automated. In our analysis, we have demonstrated that it is straightforward and inexpensive to estimate $\alpha$ and $\beta$ reliably. An alternative, which we have not addressed, would be to obtain these directly by translating the necessary calculations for each sampling strategy into floating point operations. In practice, our derivations serve well even without quantification of the initialization and recursion costs; for reasonable values of $\alpha$ and $\beta$, the acceptance probability $p_{acc}$ and the inflation factor $\nu$ can inform which of the three sampling strategies works best.

**5. Conclusion.** The prevalence of endpoint-conditioned CTMCs as an inferential tool in interdisciplinary studies has led to the development of several path-sampling algorithms. As the scope of application continues to grow, so too will the need for computationally efficient approaches, and yet this aspect has to our knowledge yet to be considered. To that end, we have presented a formal comparison of three sampling strategies: (1) modified rejection sampling, (2) direct sampling, and (3) uniformization. Significantly, we show that efficiency is a relative measure that depends heavily on the specification of the conditioned stochastic process; indeed, as demonstrated in Section 3, the computational requirements for each algorithm depend on the rate matrix $Q$, the time interval $T$, and the endpoints $a$ and $b$. We have shown that no one algorithm dominates the other two, and that each algorithm has its specific strengths and weaknesses. The previous section served to demystify those strengths and weaknesses by completely quantifying the computational costs associated with each sampling strategy.

We have concentrated our efforts on one specific application, namely, the simulation of a single sample path provided the rate matrix $Q$, the time interval $T$, and the endpoints $a$ and $b$ upon which the process is conditioned. We framed each of the three path-sampling algorithms as a progression through (1) initialization, (2) recursion, and (3) termination, and our discussion was based on an in-depth analysis of the computational requirements of each step. It should be noted that our theory is easily amenable to application-specific situations where these requirements vary; for example, it is reasonable in some cases to expect that an eigenvalue decomposition of the rate matrix has already been provided, and it is clear from the previous section how this impacts each algorithmic step.

Perhaps the most important variant to consider is the extension to the simulation of multiple sample paths. When simulating $k$ sample paths from the same endpoint-conditioned CTMC using any of the aforementioned strategies, the initialization step need only be done once. On the other hand, the iterations required for each sample path cannot in general be consolidated, and, thus, $k$ affects complexity as a scale factor of $\beta$. It follows that for large enough $k$ the initialization cost is of negligible concern, and because our examples have shown that $\alpha$ and $\beta$ are somewhat comparable, in practice, $k$ need not be that large. In such cases, complexity is determined by $k\beta E[L]$ for direct sampling and uniformization, and by $k\beta E[L]/p_{acc}$ for rejection sampling. As a result of the virtual state changes that occur when sampling by uniformization, $E[L]$ for that method will typically be somewhat larger than for the other two. For direct sampling, this is offset by a larger $\beta$, and, thus, the decision between direct sampling and uniformization rests upon the number of virtual state changes required. Rejection sampling, by contrast, completes each iteration quickly without the use of virtual transitions; it is once again the path acceptance probability that determines whether or not rejection sampling is viable. In the direct sampling algorithm we use a root finder to simulate the waiting time before the next state change. If multiple sample paths are required, it would be beneficial to completely characterize (or very accurately approximate) the cumulative distribution function for the waiting time. As soon as this task is done, drawing from the conditional waiting time distribution would be almost instantaneous. Similarly, one could, in the case of uniformization, store the values of probability masses in Remark 7 for $i \leq 3$, say. Storing the calculations allow for very fast generation of state changes from the discrete Markov chain determined by the transition probability matrix $R$. In short, the primary distinction when multiple sample paths are required is that the front-loaded procedures—direct sampling and uniformization—become comparatively more desirable, the reason being that more knowledge about the particular CTMC under consideration can be taken into account.

Finally, it should be noted that the efficiency of rejection sampling increases as the space of valid endpoint conditions is enlarged. As an example, consider the case of a CTMC observed at equidistant time points, so that the goal is to simulate sample paths using the same rate matrix $Q$ and time interval $T$ for a set of endpoint pairs $\{(a_i, b_i) : i = 1, \ldots, n\}$. In this case we can first use (unmodified) rejection sampling and assign each sample path to a pair $(a_i, b_i)$ that matches the beginning and ending state of the simulated path. If the model is appropriate, this procedure could easily account for the majority of the needed sample paths, and very few rejections would be required. The remaining sample paths can subsequently be simulated using one of the three endpoint-conditioned samplers described in this paper.

**Acknowledgments.** We thank Jeff Thorne, Ben Redelings, and Ole F. Christensen for fruitful discussions and valuable comments on the manuscript. We are very grateful to the AE and the two reviewers for many constructive comments and suggestions.

## SUPPLEMENTARY MATERIAL

**Efficient simulation from finite-state, continuous-time Markov chains with incomplete observations** (DOI: [10.1214/09-AOAS247SUPP](10.1214/09-AOAS247SUPP); .zip). We accompany our paper with R code ([www.r-project.org](www.r-project.org)) that can reproduce the figures in the manuscript [Hobolth and Stone (2009)]. A description of how the code is organized is included in the supplementary material.

DEPARTMENT OF MATHEMATICAL SCIENCES
AARHUS UNIVERSITY
DENMARK
E-MAIL: asger@daimi.au.dk

DEPARTMENT OF STATISTICS
NORTH CAROLINA STATE UNIVERSITY
RALEIGH, NORTH CAROLINA 27695
USA
E-MAIL: eric_stone@ncsu.edu